\documentclass[11pt]{article}
\usepackage{graphicx}
\def\ltap{\ \raisebox{-.4ex}{\rlap{$\sim$}} \raisebox{.4ex}{$<$}\ }
\def\gtap{\ \raisebox{-.4ex}{\rlap{$\sim$}} \raisebox{.4ex}{$>$}\ }

\begin{document} 

\noindent 
\begin{minipage}[t]{6in} 
\begin{flushright} 
TU-707           \\
hep-ph/0312228
\end{flushright} 
\end{minipage} 

\vspace{3mm}

\begin{center}

{\Large \bf 
Large $\tan\beta$ SUSY QCD corrections to $B \to X_s \gamma$ 
\footnote{
Talk at the 2nd International Conference 
on Flavor Physics (ICFP2003), KIAS, Seoul, Korea, 
Oct. 6-11, 2003} 
}    \\ 
\vspace{10mm}
{\large 
Youichi Yamada}                                 \\
\vspace{6mm}
\begin{tabular}{c}
{\it Department of Physics, Tohoku University,
          Sendai 980-8578, Japan}
\end{tabular}
\vspace{5mm}
\begin{abstract}
\baselineskip=15pt
The charged-Higgs boson contributions to the Wilson
coefficients $C_7$ and $C_8$, relevant for the decay
$B\to X_s \gamma$, are discussed in supersymmetric models 
at large $\tan \beta$.  These contributions receive 
two-loop ${\cal O}(\alpha_s\tan\!\beta)$ corrections by squark-gluino
subloops, which are possibly large and nondecoupling 
in the limit of heavy superpartners. In previous studies, 
the relevant two-loop Feynman integrals were approximated by 
using an effective two-Higgs-doublet lagrangian. However, 
this approximation is theoretically justified only 
when the typical supersymmetric scale $M_{\rm SUSY}$ is 
sufficiently larger than the electroweak scale 
$m_{\rm weak}\sim(m_W,m_t)$ and the mass of the charged-Higgs 
boson $m_{H^\pm}$. 
Here we evaluate these two-loop  integrals exactly and compare the
results with the existing, approximated ones. We then examine the
validity of this approximation beyond the region where it has been 
derived, i.e. for $m_H \gtap M_{\rm SUSY}$ and/or 
$M_{\rm SUSY}\sim m_{\rm weak}$.
\end{abstract} 
\end{center} 

\section{Introduction}
The inclusive width of the radiative decays of the $B$ mesons,
$B\to X_s \gamma$, is well described by the short-distant
processes $b\to s\gamma$ and $b \to sg$, since nonperturbative
hadronic corrections are small and well under control. The partonic
processes have been evaluated within the Standard Model (SM) 
up to the next-to-leading order in QCD~\cite{NLOSM} and 
partially beyond~\cite{NNLOproject}. 
Because in the SM the processes $b\to s\gamma$ and $b \to s g$ 
occur through loops with $W^{\pm}$ and top quark, 
possible new physics beyond the SM may contribute at 
the same level in perturbation. 
The rather good agreement between the SM prediction
and recent experimental results~\cite{EXPERIMENTS}
for the branching ratio ${\rm BR}(B\rightarrow X_s \gamma)$, 
therefore, allows already to constrain some extensions of the SM.

In the minimal supersymmetric standard model (MSSM), 
new loop contributions to the decays $b\to s \gamma$ and $b \to s g$ 
come~\cite{BSGinSUSYproposal,BSGcontributions} from the 
charged-Higgs boson $H^\pm$, 
charginos, gluino and neutralino. Their contributions are often
comparable to or even larger than the SM one. 
For generic models, these new contributions have been 
calculated~\cite{LO-generalSUSY} at the 
leading-order precision in QCD. Higher-order QCD and SUSY QCD corrections 
to these contributions 
have been evaluated~\cite{NLO-SUSY1,NLO-SUSY2a,NLO-SUSY2b} for 
specific scenarios. One important finding is that the gluino may induce 
${\cal O}(\alpha_s\tan\beta)$ corrections~\cite{NLO-SUSY2a,NLO-SUSY2b} 
to these beyond-SM contributions. 
For models with very large $\tan\beta$, the ratio of 
two Higgs VEVs, these corrections can be comparable to the leading-order 
contributions and significantly affect the 
constraints~\cite{bsglargeTB} on the charged-Higgs boson and 
SUSY particles from the experiments. 

Here we focus on the contribution of the charged-Higgs 
boson $H^\pm$ in large-$\tan\beta$ scenarios and 
analyze the two-loop ${\cal O}(\alpha_s\tan\beta)$ corrections. 
In previous 
studies~\cite{NLO-SUSY2a,NLO-SUSY2b}, squarks and gluino are assumed 
to be sufficiently heavier than the electroweak scale and 
charged-Higgs boson. 
Under this restriction, the dominant part of these corrections 
has been evaluated by using an effective two-Higgs-doublet (2HD) 
lagrangian where squarks and gluino are integrated out. 
This approach gives rather compact and simple approximated formulas 
for the ${\cal O}(\alpha_s\tan\beta)$ corrections. However, the validity 
of this approximation is not theoretically justified beyond the parameter 
range treated in Refs.~\cite{NLO-SUSY2a,NLO-SUSY2b}, i.e. 
for $m_{H^\pm}\gtap M_{\rm SUSY}$ and/or $M_{\rm SUSY}\sim m_{\rm weak}$. 
It is important to examine, in such cases, how far the 
approximation in Refs.~\cite{NLO-SUSY2a,NLO-SUSY2b} may deviate from 
the result of the exact two-loop Feynman integrals. 

In this talk, we report on the calculation of the charged-Higgs boson 
contribution to the Wilson coefficients $C_7$ and $C_8$, 
related to the processes $b \to s \gamma$ and $b\to s g$, 
to ${\cal O}(\alpha_s \tan \!\beta)$, by exact 
evaluation of the relevant two-loop diagrams.
We first present the origin of the ${\cal O}(\alpha_s \tan \!\beta)$ 
corrections to the $H^\pm$ contributions and list all necessary diagrams. 
We next review the approximation in Refs.~\cite{NLO-SUSY2a,NLO-SUSY2b}, 
here called the nondecoupling approximation. 
Finally we make a numerical comparison of the exact result and 
the nondecoupling approximation of the ${\cal O}(\alpha_s \tan \beta)$ 
results for the $H^\pm$ contributions to $C_7(\mu_W)$ and $C_8(\mu_W)$, 
and discuss the validity of the approximation. 
A more complete discussion is presented in Refs.~\cite{BGY,fullpaper}. 

\section{${\cal O}(\alpha_s\tan\!\beta)$ corrections to the 
$H^\pm$ contribution}
\label{diagrams}

\begin{figure}[h] 
\vspace{0.3truecm}
\begin{center} 
\includegraphics[width= 6.5cm]{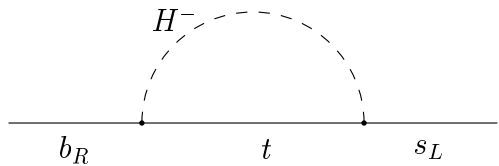}
\end{center} 
\caption[f1]{\small 
$b\to s\gamma$ and $b\to sg$ by the one-loop charged-Higgs boson exchange. 
The photon or gluon is to be attached at the $t$ or $H^-$ 
lines.}
\label{CH-0}
\end{figure}

In the MSSM with large $\tan\beta$, the dominant part 
of the one-loop $H^{\pm}$ contributions to the 
$b\to s \gamma$ and $b \to s g$ decays comes from the 
diagrams in Fig.~\ref{CH-0}, where the 
photon or gluon is to be attached to the $t$ or the $H^\pm$ lines. 
The enhancing factor $\tan\beta$ at the $\bar{t}_Lb_RH^+$ vertex is 
cancelled by the suppressing factor $\cot\beta$ at the $\bar{s}_Lt_RH^-$ 
vertex. 

\begin{figure}[t] 
\begin{center} 
\includegraphics[width= 6.5cm]{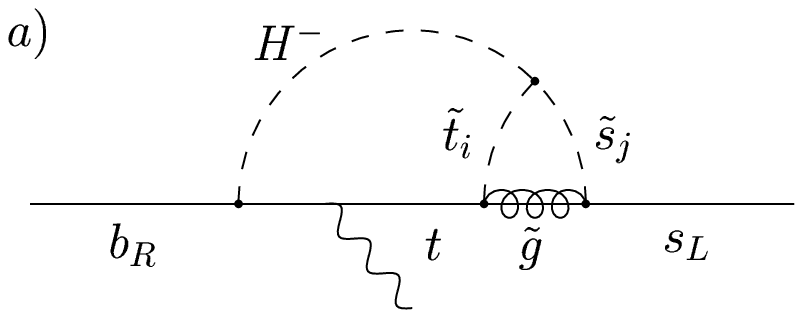}
\hspace{5mm}
\includegraphics[width= 6.5cm]{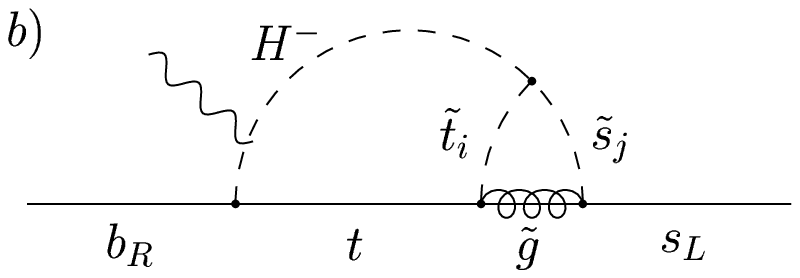}
\end{center} 
\vspace{-0.1truecm}
\label{CH-phHtop}
%
\begin{center} 
\includegraphics[width= 6.5cm]{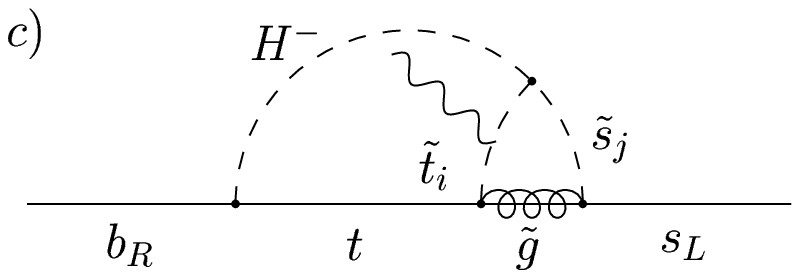}
\hspace{5mm}
\includegraphics[width= 6.5cm]{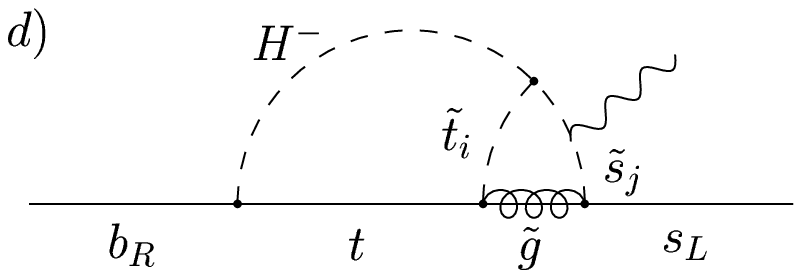}
\end{center} 
\vspace{-0.2truecm}
%
\begin{center} 
\includegraphics[width= 6.5cm]{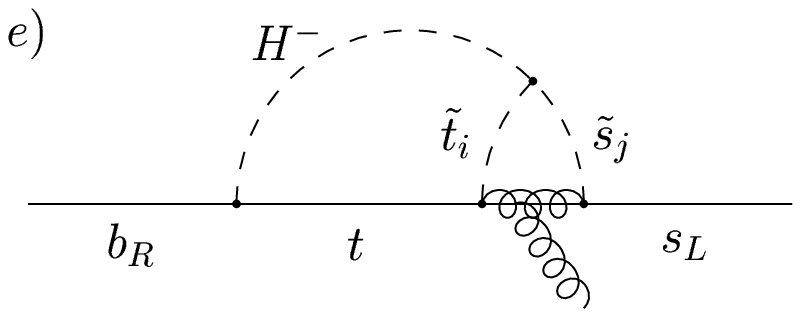}
\end{center} 
\vspace{-0.2truecm}
\caption[f1]{\small $H^\pm$ mediated diagrams contributing 
 at order ${\cal O}(\alpha_s \tan \beta)$ to the decays 
 $b\to s\gamma$ and $b\to s g$. The photon must be replaced by a 
 gluon and vice versa, whenever possible.} 
\label{CHexch2loops}
\end{figure}

It has been shown \cite{NLO-SUSY2a,NLO-SUSY2b} that 
these $H^\pm$ contributions receive 
${\cal O}(\alpha_s\tan\!\beta)$ corrections from the squark-gluino 
subloops, which are potentially large for large $\tan\beta$. 
These corrections may arise from 
\begin{enumerate}
\item 
counterterm for the $H^+\bar{t}_Lb_R$ coupling~\cite{dmb,carenaH0,carenaHp}, 
coming from the mass corrections $\delta m_b$; 
\item 
proper vertex corrections to the 
$H^-\bar{s}_Lt_R$ coupling~\cite{NLO-SUSY2a,NLO-SUSY2b};
\item 
effective four-point couplings 
$H^-\bar{s}t\gamma$ and  $H^-\bar{s}tg$, 
as well as the $H^-\bar{s}_Lt_L$ coupling, 
generated by squark-gluino subloops. 
\end{enumerate}
The two-loop diagrams relevant to the corrections of type 2 and 3, 
listed above, are shown in Fig.~\ref{CHexch2loops}, where 
the photon must be replaced by a gluon, and vice versa, 
whenever possible. 
Note that, while the one-loop diagram in Fig.~\ref{CH-0} has an 
chirality flip in the internal top quark line, the 
diagrams in Fig.~\ref{CHexch2loops} can have such a chirality flip also
on the $\tilde{t}$-squark line, giving rise to the 
effective $H^-\bar{s}_Lt_L$ coupling mentioned above. 
Both eigenstates of 
the $\tilde{t}$-squark, $\tilde{t}_1$ and $\tilde{t}_2$, 
but only one $\tilde{s}$-squark eigenstate, the left-handed one, 
contribute in Fig.~\ref{CHexch2loops}. 

We calculate the $H^{\pm}$ contributions to the Wilson 
coefficients $C_7$ and $C_8$ at the electroweak scale $\mu_W=m_W$, 
including the ${\cal O}(\alpha_s\tan\beta)$ corrections. 
Our normalization of $C_7(\mu_W)$ and $C_8(\mu_W)$ 
is the conventional one, as 
follows from the definition of the effective Hamiltonian, 
\begin{equation}
H_{\rm eff} \supset
 -\frac{4G_F}{\sqrt{2}}V^*_{ts}V_{tb}
 \left( C_7(\mu) {\cal O}_7(\mu) + C_8(\mu) {\cal O}_8(\mu) 
 \right) \,.
\end{equation}
and of the operators ${\cal O}_7$ and ${\cal O}_8$, 
\begin{equation}
{\cal O}_7(\mu) = 
 \frac{e}{16\pi^2} m_b(\mu)\bar{s}_L\sigma^{\mu\nu} b_R F_{\mu\nu}\,, 
\hspace{0.8truecm}
{\cal O}_8(\mu) = 
 \frac{g_s}{16\pi^2}m_b(\mu)\bar{s}_L\sigma^{\mu\nu}T^a b_RG^a_{\mu\nu}\,, 
\label{O7and8}
\end{equation}
where $F_{\mu\nu}$ and $G^a_{\mu\nu}$ are 
the field strengths of the photon and the gluon, respectively.  

We denote by 
$ C_{7,H}(\mu_W)$ and $ C_{8,H}(\mu_W)$ the $\tan \beta$-unsuppressed
$H^\pm$ contribution to $C_7(\mu_W)$ and $C_8(\mu_W)$, respectively. 
They are decomposed as
\begin{equation}
C_{i,H}(\mu_W) = \frac{1}{1+\Delta_{b_R,b} \tan\beta}
\left[ C_{i,H}^0(\mu_W) + \Delta C_{i,H}^1(\mu_W) \right] \,,
\label{defWC}
\end{equation}
where $C_{i,H}^0(\mu_W)$ and $\Delta C_{i,H}^1(\mu_W)$
are the contributions of the one-loop diagram in Fig.~\ref{CH-0} 
and the two-loop diagrams in Fig.~\ref{CHexch2loops}, respectively. 
The overall factor $1/(1+\Delta_{b_R,b} \tan\beta)$ represents 
the correction to the $H^+\bar{t}_L b_R$ Yukawa coupling 
coming from the correction to $m_b$~\cite{dmb,carenaH0}. 
The one-loop function $\Delta_{b_R,b}$ is given in Ref.~\cite{fullpaper} 
and of the order of 
$\alpha_s\mu m_{\tilde{g}}/M_{\rm SUSY}^2\sim \alpha_s M_{\rm SUSY}^0$. 
In the large $M_{\rm SUSY}$ limit, 
the contributions from $\Delta_{b_R,b}$ and the vertex corrections 
to the $H^-\bar{s}_L t_R$ coupling are nondecoupling, 
while all other contributions of 
Fig.~\ref{CHexch2loops} are decoupling.

We calculate all contributions of the two-loop 
diagrams in Fig.~\ref{CHexch2loops} by exact evaluation of 
the loop integrals, making use of 
results and techniques in Ref.~\cite{GvdB}. 
The explicit forms of the Feynman integrals for these 
diagrams are listed in Ref.~\cite{fullpaper}.

\section{Nondecoupling approximation vs. exact calculation}
\label{EFLvsHME}
In Refs.~\cite{NLO-SUSY2a,NLO-SUSY2b}, the calculations of the 
${\cal O}(\alpha_s \tan\beta)$ corrections were performed under the 
assumption that all squarks and gluino, around $M_{\rm SUSY}$, 
are sufficiently heavier than the top quark and the $W$ boson, 
whereas $m_{H^\pm}$ is around the electroweak scale 
$m_{\rm weak}\sim m_W, m_t$. The squark-gluino subloop 
corrections to the $H^\pm$ contributions were described
in terms of an effective 2HD lagrangian, 
in which squarks and gluino are integrated out. 
In the following we call the approximation in 
Refs.~\cite{NLO-SUSY2a,NLO-SUSY2b} the 
nondecoupling approximation, since it collects all 
nondecoupling parts of the 
${\cal O}(\alpha_s \tan\beta)$ corrections. 
Strictly speaking, however, it includes parts of 
the formally decoupling ${\cal O}(m_{\rm weak}^2/M_{\rm SUSY}^2)$ 
contributions through the masses and couplings of squarks~\cite{OTHERPAPS}. 
For the contribution from $\delta m_b$, the use of the effective 
2HD lagrangian allows us to resum higher-order 
${\cal O}((\alpha_s \tan\beta)^n)$ terms~\cite{carenaHp} in 
Eq.~(\ref{defWC}), by putting $\Delta_{b_R,b}$ in the denominator. 

For $\Delta C_{i,H}^1(\mu_W)$ coming from the 
diagrams in Fig.~\ref{CHexch2loops}, the nondecoupling 
approximation is obtained by retaining only the diagrams a) and b), 
with chirality flip on the $t$-quark line only, and evaluating the 
squark-gluino subloops at vanishing external momenta. 
By this approximation, the original two-loop Feynman integrals 
for the ${\cal O}(\alpha_s \tan\beta)$ corrections are 
factorized into two one-loop diagrams, taking rather compact forms. 

We show one example to illustrate the difference 
between the nondecoupling approximation and the exact calculation. 
The contribution of the diagram Fig.~\ref{CHexch2loops}a), 
with chirality flip on the $t$-quark line, is proportional to the integral 
\begin{eqnarray}
I_{ti2}  
& = &
\mu m_{\tilde{g}} 
\int\frac{d^4 k}{(2 \pi)^4} \
\int\frac{d^4 l}{(2 \pi)^4} \
\frac{1}{\left[ k^2    -m_t^2            \right]^2 
         \left[ k^2    -m_{H^\pm}^2            \right]
         \left[ (l+k)^2-m_{\tilde{t}_i}^2\right]
         \left[ l^2    -m_{\tilde{s}}^2  \right] 
         \left[ l^2    -m_{\tilde{g}}^2  \right]} 
\nonumber\\ && 
\times 
\left\{ \frac{l\cdot k}{l^2 -m_{\tilde{g}}^2} 
       -\frac{2k^2}{k^2 -m_t^2} \right\}\, .
\label{intt2}
\end{eqnarray}
The loop momenta $k$ and $l$ represent the momenta of $(t, H^{\pm})$ and 
SUSY particles, respectively. 
In the nondecoupling approximation, the $k$-dependence of the 
$\tilde{t}_i$ line is neglected, i.e. $(l+k)^2-m_{\tilde{t}_i}^2$ is 
replaced by $l^2-m_{\tilde{t}_i}^2$. 
The term proportional to $l\cdot k$ is then dropped and 
the integral is factorized into two one-loop integrals as 
\begin{equation}
I_{ti2}\vert_{\rm nondec}
= 
\mu m_{\tilde{g}} 
\int\frac{d^4 k}{(2 \pi)^4} \
\frac{-2k^2}{\left[ k^2 -m_t^2 \right]^3
         \left[ k^2    -m_{H^\pm}^2     \right] }
\int\frac{d^4 l}{(2 \pi)^4} \
\frac{1}{
         \left[ l^2-m_{\tilde{t}_i}^2\right]
         \left[ l^2    -m_{\tilde{s}}^2  \right]
         \left[ l^2    -m_{\tilde{g}}^2  \right]} \,.
\label{eq-ti2-nondec}
\end{equation}

In the nondecoupling approximation, 
${\cal O}((m_{\rm weak}^2,m_{H^\pm}^2)/M_{\rm SUSY}^2)$ terms which may 
come from the $k$-dependence of the squark-gluino subloops of 
the diagrams in Fig.~\ref{CHexch2loops}a,b), as well as the whole 
contributions of the diagrams in Fig.~\ref{CHexch2loops}c-e), 
are neglected. 
The resulting deviation of this approximation from 
the exact two-loop calculation is, therefore, expected to 
become large when $M_{\rm SUSY}$ is not much 
heavier than $m_{\rm weak}$ and/or $m_{H^\pm}$. 

Since the condition for the theoretical justification of the 
nondecoupling approximation, 
$m_{\rm weak}^2\sim m_{H^\pm}^2\ll M_{\rm SUSY}^2$, is 
often violated in well-known scenarios for the SUSY breaking mechanism, 
it is very important to study how far this approximation may be 
applied beyond this restricted parameter region. 
Clearly, a definite answer to this question will be given by 
the exact calculation of all the two-loop 
diagrams in Fig.~\ref{CHexch2loops}, without 
any assumption on the relative size of $m_{H^\pm}$, $M_{\rm SUSY}$, 
and $m_{\rm weak}$.

\section{Numerical results} 
\label{WCoeff}

We present numerical results for the $H^\pm$ contributions, 
$C_{i,H}(\mu_W)(i=7,8)$ shown in Eq.~(\ref{defWC}), 
at the scale $\mu_W=M_W$. 
We make a comparison of the results 
$C_{i,H}(\mu_W)\vert_{\rm exact}$ obtained from the exact 
two-loop integrals with the nondecoupling approximation 
$C_{i,H}(\mu_W)\vert_{\rm nondec}$. 

\begin{figure}[h] 
\vspace{0.3truecm}
\begin{center} 
\includegraphics[width= 7.8cm]{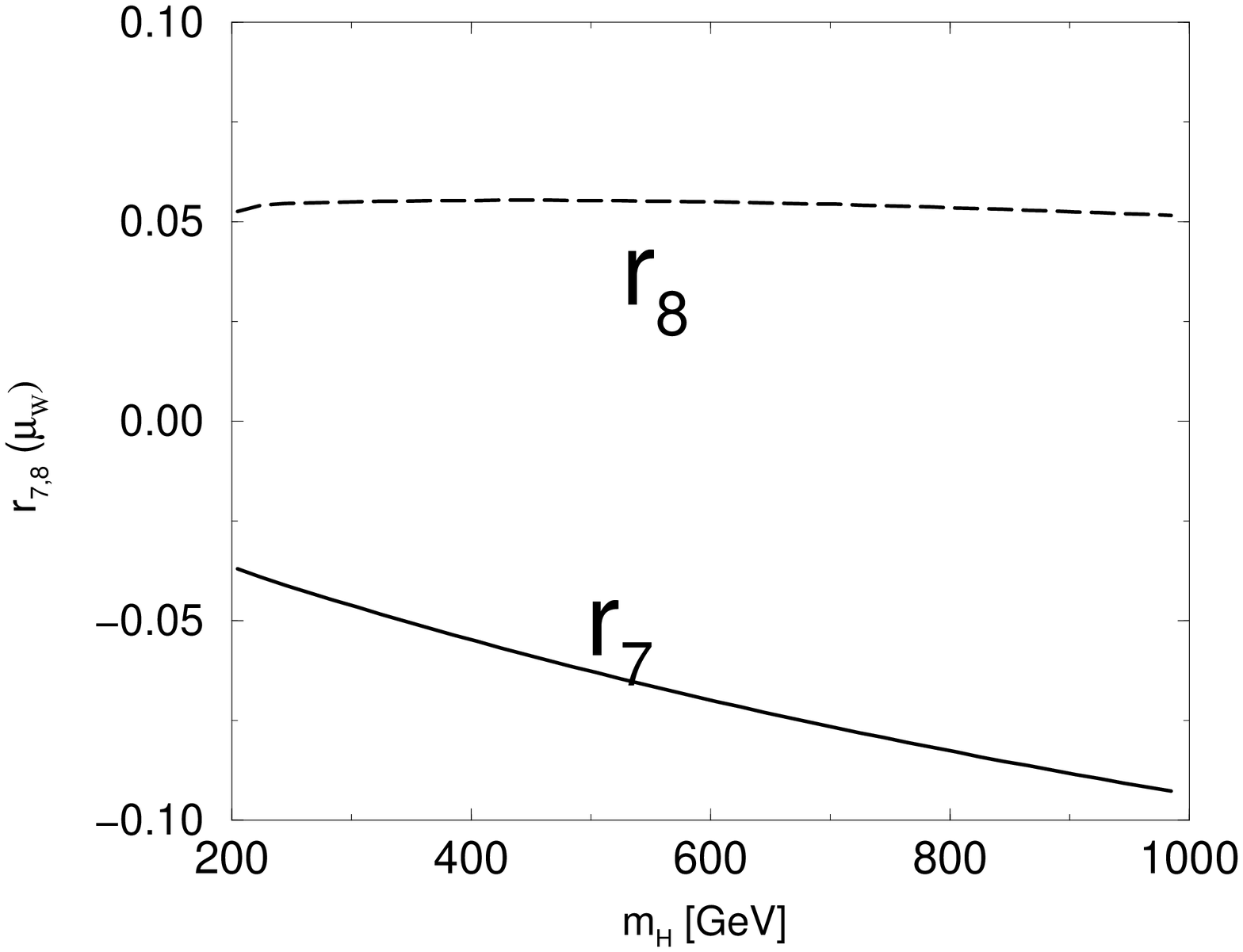}
\hspace{3mm}
\includegraphics[width= 7.8cm]{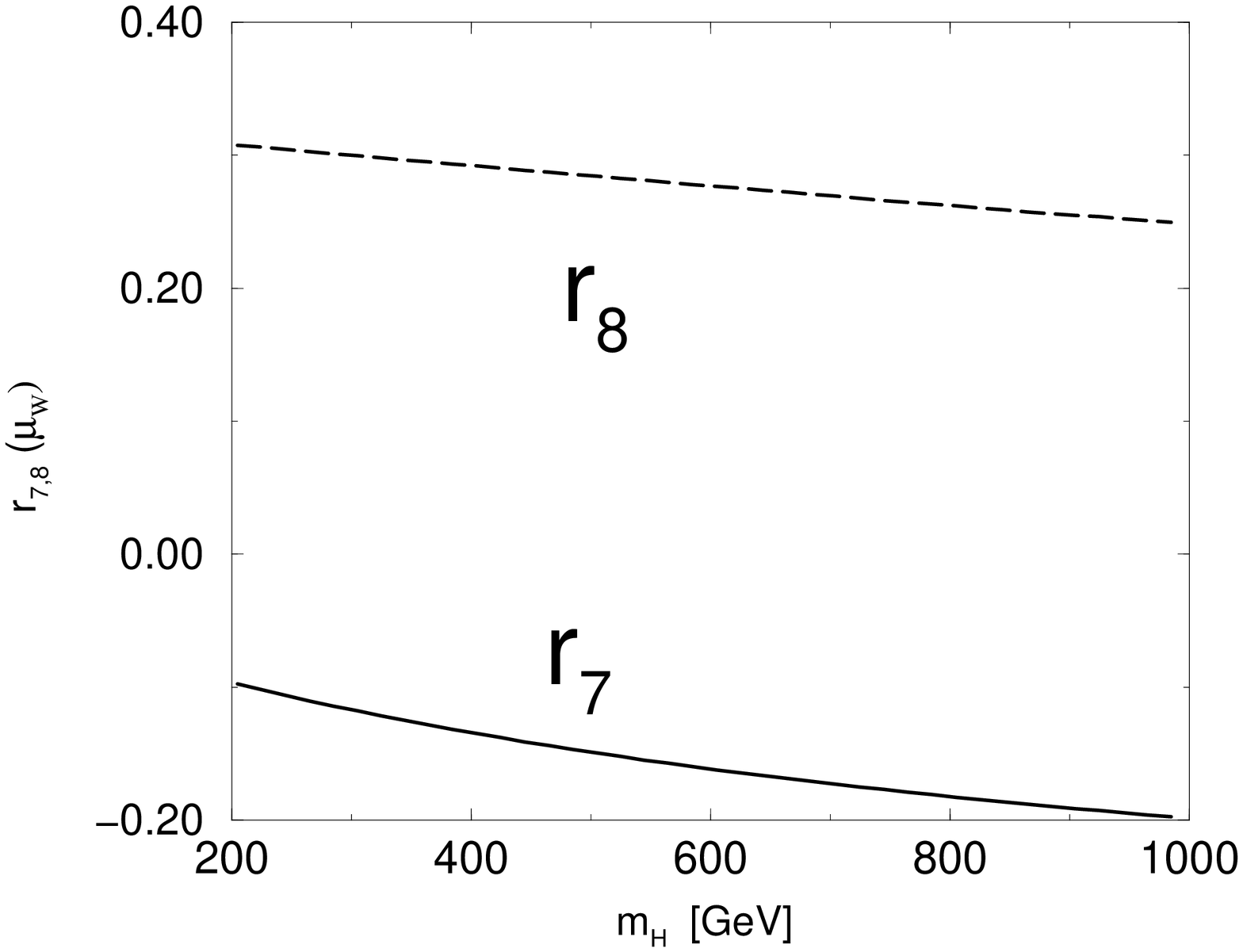} 
\end{center} 
\vspace{-0.5truecm}
\caption[f1]{\small Relative deviations $r_i(\mu_W)(i=7,8)$ of 
 the nondecoupling approximations of the ${\cal O}(\alpha_s \tan\beta)$ 
 Wilson coefficients $C_{i,H}(\mu_W)$ from the exact two-loop results, 
 as defined in the text, 
 for the spectrum I (left) and spectrum II (right).
}
\label{r78fig} 
\end{figure} 
In Fig.~\ref{r78fig}, we plot the ratios
\begin{equation}
r_i(\mu_W) \equiv 
\displaystyle{
  \frac{C_{i,H}(\mu_W)\vert_{\rm nondec} -C_{i,H}(\mu_W)\vert_{\rm exact}}
       {C_{i,H}(\mu_W)\vert_{\rm exact}} } 
\hspace*{5truemm}
 (i = 7,8)\,,
\label{plotratio}
\end{equation}
showing the relative deviation of the nondecoupling approximation from 
the exact two-loop calculation, as functions of $m_{H^\pm}$. 
The correction $\Delta_{b_R,b}$ in Eq.~(\ref{defWC}), 
coming from the mass correction $\delta m_b$, 
cancel out in the ratios $r_i$. 

Two sets of the parameters for squarks and gluino are used in 
Fig.~\ref{r78fig}. For an example of a heavier SUSY 
spectrum, called here spectrum~I, we have chosen 
$(m_{\tilde{s}_L},m_{\tilde{t}_1},m_{\tilde{t}_2})=
 (700,500,450)$ GeV, the left-right mixing angle of $\tilde{t}$-squarks 
$\cos \theta_t = 0.8$, $m_{\tilde{g}}=600$ GeV, and $\mu=550$ GeV. 
For a lighter SUSY spectrum, spectrum~II, we set 
$(m_{\tilde{s}_L},m_{\tilde{t}_1},m_{\tilde{t}_2})=
 (350,400,320)$ GeV, $\cos \theta_t = 0.8$, 
$m_{\tilde{g}}=300$ GeV, and $\mu=450$ GeV. 
As for other input parameters, we have used 
$\tan\!\beta=30$, 
$m_t(\mu_W) = 176.5\,$GeV, which corresponds to a pole mass 
$M_t=175\,$GeV, and $\alpha_s(\mu_W) =0.12 $.

For the spectrum I, the difference between 
the exact calculation and the nondecoupling approximation is 
very small in the whole range of $m_{H^{\pm}}$, 
even for $m_{H^{\pm}}\gtap M_{\rm SUSY}$. 
This is a surprising result since, as
discussed in Sect.~\ref{EFLvsHME}, 
the ${\cal O}(m_{H^\pm}^2/M_{\rm SUSY}^2)$ deviation 
was expected to be large in this region. 
In the case of the spectrum~II, $r_{7,8}$ become larger. 
The corrections beyond the nondecoupling approximation are of the
same order of the SU(2)$\times$U(1) breaking effects in the SUSY
particle subloops~\cite{OTHERPAPS} and are no longer negligible.
Nevertheless, $r_7$ and $r_8$ 
remain of the same order of magnitude for increasing
$m_{H^{\pm}}$, up to $m_{H^{\pm}}\gg M_{\rm SUSY}$.
In both cases, the main part of the difference between the 
result of the nondecoupling approximation and the exact two-loop result
comes from the diagram in Fig.~\ref{CHexch2loops}a) and, for $C_{8,H}$, 
also from the diagram in Fig.~\ref{CHexch2loops}e).

To understand the results for $m_{H^\pm}\gtap M_{\rm SUSY}$
qualitatively, we again consider the diagram in Fig.~\ref{CHexch2loops}a),
with chirality flip on the top quark line. When 
$m_{H^\pm}$ is sufficiently larger than $m_t$,
this diagram gives the largest contribution to $\Delta C_{7,H}^1(\mu_W)$ and 
$\Delta C_{8,H}^1(\mu_W)$. 
It is proportional to the integral $I_{ti2}$ in Eq.~(\ref{intt2}). 
For the following discussion
we rewrite  $I_{ti2}$ in the form
\begin{equation}
 I_{ti2}(m_t,m_{H^\pm}, m_{\tilde{t}_i},m_{\tilde{s}},m_{\tilde{g}}) =
\int\frac{d^4 k}{(2 \pi)^4} \
\frac{k^2}
{\left[k^2 -m_t^2\right]^3 \left[k^2 - m_{H^\pm}^2\right]}
 \,
 Y_{ti2}\left(k^2;m_{\tilde{t}_i},m_{\tilde{s}},m_{\tilde{g}}
       \right)\,.  
\label{eq-Iint}
\end{equation}
$Y_{ti2}(k^2;m_{\tilde{t}_i},m_{\tilde{s}},m_{\tilde{g}})$
represents the form factor for the effective vertex $H^-\bar{s}_Lt_R$ 
generated by the squark-gluino loops and is given by: 
\begin{equation}
Y_{ti2}(k^2;m_{\tilde{t}_i},m_{\tilde{s}},m_{\tilde{g}}) =
\mu m_{\tilde{g}} 
 \left[-2 F + (k^2-m_t^2) G\right]
\left(k^2;m_{\tilde{t}_i}^2,m_{\tilde{s}}^2,m_{\tilde{g}}^2\right) \,,
\label{eq-Ydef} 
\end{equation}
with 
\begin{eqnarray}
F(k^2; m_{\tilde{t}_i}^2,m_{\tilde{s}}^2,m_{\tilde{g}}^2)
 & = & 
\int\frac{d^4 l}{(2 \pi)^4} \
\frac{1}{ \left[ (l+k)^2 - m_{\tilde{t}_i}^2 \right]
\left[ l^2 -\!m_{\tilde{s}}^2 \right]
\left[ l^2-m_{\tilde{g}}^2 \right] } \,,
\label{eq-Fdef}
\\
k^{\mu} 
G(k^2; m_{\tilde{t}_i}^2,m_{\tilde{s}}^2,m_{\tilde{g}}^2) 
 & = & 
\int\frac{d^4 l}{(2 \pi)^4} \
\frac{l^{\mu}}{ \left[ (l+k)^2 - m_{\tilde{t}_i}^2 \right]
\left[ l^2 -\!m_{\tilde{s}}^2 \right]
\left[ l^2-m_{\tilde{g}}^2 \right]^2} \,.
\label{eq-Gdef}
\end{eqnarray}
The nondecoupling approximation of $I_{ti2}$, shown in 
Eq.~(\ref{eq-ti2-nondec}), is obtained 
by replacing $Y_{ti2}$ in Eq.~(\ref{eq-Iint}) with 
\begin{equation}
 Y_{ti2}\vert_{\rm nondec} =
 - 2 \mu m_{\tilde{g}} F (0; m_{\tilde{t}_i}^2,m_{\tilde{s}}^2,
m_{\tilde{g}}^2)\,,
\label{eq-Yapprox}
\end{equation}
which is an ${\cal O}(M_{\rm SUSY}^0)$ constant with respect to $k^2$. 
To simplify our discussion, 
we set hereafter ($m_{\tilde{t}_i}$, $m_{\tilde{s}}$, $m_{\tilde{g}}$, $\mu$) 
equal to $M_{\rm SUSY}$.

For $\vert k^2 \vert \ll M_{\rm SUSY}^2$, $F(k^2;M_{\rm SUSY}^2)$ 
and $G(k^2;M_{\rm SUSY}^2)$ behave as 
\begin{eqnarray}
F(k^2; M_{\rm SUSY}^2) 
& = &
 {\cal O}\left(\frac{1}{M_{\rm SUSY}^{2}}\right) + 
 {\cal O}\left(\frac{k^2}{M_{\rm SUSY}^{4}}\right)\,, 
\nonumber \\
G(k^2; M_{\rm SUSY}^2) 
& = &
 {\cal O}\left(\frac{1}{M_{\rm SUSY}^{4}}\right)\,.
\label{eq-FHzeroKlimit}
\end{eqnarray} 

For $\vert k^2 \vert \gg M_{\rm SUSY}^2$, it is:
\begin{eqnarray}
F(k^2; M_{\rm SUSY}^2) 
& \to & 
{\cal O}\left( \frac{1}{k^2} \ln \frac{k^2}{M^2_{\rm SUSY}} \right), 
\nonumber \\
G(k^2; M_{\rm SUSY}^{2})  
& \to & 
{\cal O}\left( \frac{1}{k^4}\ln \frac{k^2}{M_{\rm SUSY}^2} \right) \,.
\end{eqnarray}
The behavior of $Y_{ti2}(k^2; M_{\rm SUSY}^2)$ is therefore 
\begin{equation}
Y_{ti2}(k^2; M_{\rm SUSY}^2)  \to  
\left\{ 
\begin{array}{ll} 
Y_{ti2}\vert_{\rm nondec} + 
 {\cal O}\left(
 \displaystyle{\frac{k^2}{M_{\rm SUSY}^2}},
 \displaystyle{\frac{m_t^2}{M_{\rm SUSY}^2}} \right)  
& 
(\vert k^2 \vert \ll M_{\rm SUSY}^2), 
\\[1.8ex] 
{\cal O}\left(\displaystyle{\frac{M_{\rm SUSY}^2}{k^2} 
 \ln \frac{k^2}{M_{\rm SUSY}^2}} \right) 
& 
(\vert k^2 \vert \gg M_{\rm SUSY}^2)\,, 
\end{array} 
\right. 
\label{eq-deviation}
\end{equation}
which supports the naive expectation that a substantial deviation 
of $I_{ti2}(m_t,m_{H^\pm},M_{\rm SUSY}^2)$ from its nondecoupling 
approximation $I_{ti2}(m_t,m_{H^\pm},M_{\rm SUSY}^2)\vert_{\rm nondec}$ may 
arise for $m_{H^\pm}\gtap M_{\rm SUSY}$. 

However, the factor multiplying $Y_{ti2}(k^2; M_{\rm SUSY}^{2})$ in
Eqs.~(\ref{eq-Iint}) plays an important role, leading 
to the fact that
this expectation does not hold in the case in which $M_{\rm SUSY}$ 
is not rather light. 
Since for $\vert k^2\vert \gg m_{H^\pm}^2$ this factor drops as
$d^4k/k^6$, the integral~(\ref{eq-Iint}) gets its largest contribution
from the region $\vert k^2\vert \ltap m_{H^\pm}^2$.  A closer
inspection actually shows that it is the region of small 
$\vert k^2\vert$, up to $\vert k^2\vert = {\cal O}( m_t^2)$, which 
determines the bulk of the value of this integral. If $M_{\rm SUSY}$ is 
sufficiently larger than $m_t$, 
$Y_{ti2}(k^2; M_{\rm SUSY}^2)$ does not
deviate substantially from $Y_{ti2}\vert_{\rm nondec}$ in this region. 
This explains the smallness of the deviation for 
$m_{H^\pm}\gtap M_{\rm SUSY}$ shown in Fig.~\ref{r78fig}. 

\section{Conclusion}
\label{conclu}
We have studied the ${\cal O}(\alpha_s \tan \beta)$
corrections to the $H^{\pm}$ contributions to the Wilson
coefficients relevant for the decay $B\to X_s \gamma $, in the 
MSSM with large $\tan\beta$.  These corrections are
generated by the shift of the $b$-quark mass in the Higgs-quark
couplings and by the dressing of the one-loop $H^\pm$ diagrams
with squark-gluino subloops, as shown in Fig.~\ref{CHexch2loops}.  

In this talk, we have focused on the latter class of corrections. 
In previous studies~\cite{NLO-SUSY2a,NLO-SUSY2b}, the contributions 
from these two-loop diagrams were calculated in an approximated 
way, by using an effective 2HD lagrangian formalism in which squarks and
gluino are integrated out.  This method, here called the 
nondecoupling approximation, is theoretically justified 
in the case $m_{\rm weak}^2\sim m_{H^\pm}^2 \ll M_{\rm SUSY}^2$, 
and gives rather compact forms for these corrections. 
However, the deviation from the exact two-loop result was, 
in principle, expected to be of 
${\cal O}(m_{\rm weak}^2,m_{H^\pm}^2/M_{\rm SUSY}^2)$, and to become 
significant when $m_{\rm weak} \le M_{\rm SUSY} $ and/or 
$m_{H^\pm} \ge M_{\rm SUSY}$.

We have calculated the contributions of the two-loop diagrams 
in Fig.~\ref{CHexch2loops} exactly, without assuming any patterns for 
the mass of the particle involved, and compared the results with the 
nondecoupling approximation. 
Surprisingly, the difference between the nondecoupling approximation in 
Refs.~\cite{NLO-SUSY2a,NLO-SUSY2b} and the 
exact two-loop result was shown to be quite small, even for 
$m_{H^\pm} \gtap M_{\rm SUSY}$, provided $M_{\rm SUSY}$ is sufficiently 
larger than $m_{\rm weak}$. The unexpected absence of 
large deviation for the case of $m_{H^\pm} \gtap M_{\rm SUSY}$ 
with $M_{\rm SUSY}^2\gg m_{\rm weak}^2$ can be 
understood from the structure of the relevant two-loop integrals. 
In contrast, nonnegligible deviation appeared 
for $M_{\rm SUSY}$ not much larger than $m_{\rm weak}$. 

We have illustrated our findings by showing the 
$H^\pm$ contributions to the Wilson coefficients $C_{7,H}$ 
and $C_{8,H}$ at the electroweak matching scale $\mu_W$, 
for different spectra of the gluino, squarks and $H^\pm$. 
Analyses of $C_7$ and $C_8$ at a low scale $\sim m_b$, 
including other contributions than the $H^\pm$-mediated one, 
and of the actual branching ratio ${\rm BR}(B \to X_s \gamma)$, 
will be presented in future work.

\vspace*{0.5truecm}
\noindent 
{\bf Acknowledgements}  
This talk is based on the works~\cite{BGY,fullpaper} in collaboration 
with Francesca Borzumati and Christoph Greub. 
The author was supported by the 
Grant-in-aid for Scientific Research from the Ministry of Education,
Culture, Sports, Science, and Technology of Japan, No.~14740144.

\end{document}